\documentclass[reprint,superscriptaddress]{revtex4-1}

\usepackage{graphicx}
\usepackage{subfigure}
\usepackage{color}
\usepackage{amsmath}
\usepackage{amssymb}
\usepackage{epstopdf}
\usepackage{physics}
\usepackage[linktocpage,colorlinks=true,linkcolor=blue,citecolor=blue,breaklinks=true]{hyperref}
\usepackage{verbatim}
\usepackage{breakcites}
\usepackage[version=3]{mhchem}

\newcommand{\be}{\begin{equation}}
\newcommand{\ee}{\end{equation}}

\begin{document}

\preprint{}

\title{A stochastic model for the turbulent ocean heat flux under Arctic sea ice}

\author{S. Toppaladoddi}
\affiliation{School of Mathematics, University of Leeds, Leeds LS2 9JT, U.K.}
\affiliation{All Souls College, Oxford OX1 4AL, U.K.}
\email[]{S.Toppaladoddi@leeds.ac.uk}

\author{A. J. Wells}
\affiliation{Atmospheric, Oceanic and Planetary Physics, Department of Physics, University of Oxford, Oxford OX1 3PU, U.K.}

\date{\today}

\begin{abstract}
The physics of planetary climate features a variety of complex systems that are challenging to model as they feature turbulent flows. A key example is the heat flux from the upper ocean to the underside of sea ice which provides a key contribution to the evolution of the Arctic sea ice cover.  Here, we develop a model of the turbulent ice-ocean heat flux using coupled ordinary stochastic differential equations to model fluctuations in the vertical velocity and temperature in the Arctic mixed layer. All the parameters in the model are determined from observational data. A detailed comparison between the model results and measurements made during the Surface Heat Budget of the Arctic Ocean (SHEBA) project reveals that the model is able to capture the probability density functions (PDFs) of velocity, temperature and heat flux fluctuations. Furthermore, we show that the temperature in the upper layer of the Arctic ocean can be treated as a passive scalar during the whole year of SHEBA measurements. The stochastic model developed here provides a computationally inexpensive way to compute an observationally consistent PDF of this heat flux, and has implications for its parametrization in regional and global climate models.

\end{abstract}


\maketitle

\section{Introduction}
Many systems of interest in both the natural and engineered environments possess a very large number of degrees of freedom, which makes the use of statistical physics the only feasible way to study their dynamics \citep{Landau-statphys}. These systems -- either in a state of equilibrium or disequilibrium -- display fluctuations in key physical quantities that describe them, and a complete description of the systems must include these fluctuations \citep{chandra1943}. A systematic way to study and model these fluctuations is provided by mathematical tools which fall under the rubric of stochastic methods \citep{chandra1943, wang1945, gardiner1985_book}, which have been fruitfully used to study fluctuations in systems such as the motion of a Brownian particle \citep{chandra1943}, chemical reactions \citep{van1992}, turbulent flows \citep{Pope}, and the Earth's climate \citep{Saltzman:2002, Dijkstra}. Here, we use stochastic methods to study the fluctuations in the turbulent ocean heat flux (hereafter referred to as ocean heat flux for short) under Arctic sea ice, which contributes to the melting of the ice cover and is an under-constrained element in the description of the Arctic climate system \citep{MU71}.

Arctic sea ice is one of the most sensitive components of the Earth's climate system, and plays an important role in the Earth's radiation budget due to its high albedo. The evolution of the ice cover is affected by processes that act on disparate length and time scales -- from the transport of salt, momentum, and heat in the boundary layers next to the ice-ocean interface to the atmospheric drivers of ice motion including ice export through the Fram Strait. These processes originate from the nonlinear interactions between sea ice and the atmosphere and underlying ocean \citep{OneWatt}. Accurate modelling of these interactions is both challenging and indispensable for reliable predictions of the fate of Arctic sea ice \citep{perovich2009}. 


The principal source of the ocean heat flux is the shortwave radiation absorbed by the upper layers of the ocean during summer \citep{maykut1995, stanton2012}. Whilst some of this heat is turbulently mixed to melt the neighbouring ice, the creation of cold and relatively fresh water due to ice ablation traps some of this absorbed heat at depth creating a near-surface of maximum temperature beneath the mixed layer \citep{timmermans2015}. The  latter heat store is released to the ice-ocean interface when fluid motions due to shear and buoyancy ensue in fall and winter \citep{timmermans2015}. Upward fluxes of heat from the deeper ocean provide an additional, but comparatively insignificant, contribution \citep{timmermans2008}. 

The role of the ocean heat flux in the growth of sea ice was systematically studied by \citet{MU71}. In their observationally consistent one-dimensional thermodynamic model, the ocean heat flux was used as an input parameter due to scarcity of measurements \citep{MU71}. Their results showed that a variation in the heat flux from 0 to 7 Wm$^{-2}$ changed the mean thickness of sea ice from about 6 m to 0 m, thus highlighting the importance of this heat flux and the necessity of measuring and accurately modelling it.

Since the pioneering oceanic boundary layer measurements of \citet{mcphee1976} during the Arctic Ice Joint Dynamics Experiment, there have been several experimental studies that have either directly \citep{moum1990, mcphee1992, perovich2002, shaw2009, stanton2012, peterson2017} or indirectly \citep{mcphee1982, JSW91} measured the ocean heat flux in the Arctic. The time series analysis by \citet{mcphee1992} has revealed the following interesting features regarding the ocean heat flux: (i) a large fraction of heat transport occurs in intermittent bursts which can be larger than the mean heat flux by an order of magnitude; (ii) the probability density functions (PDFs) of the ocean heat flux have large values of skewness and kurtosis, and hence are not Gaussian; and (iii) the PDFs obtained from heat flux time series on different days and at different depths have a self-similar form, which can be fit with separate stretched exponential functions for the positive and negative segments. However, significant intersite differences in the heat flux have been observed and attributed to local topographical features of the ice-ocean interface \citep{JSW91}. This is because the interfacial roughness enhances the interaction between flows inside and outside the boundary layers, leading to an increased heat flux when compared to that of a flat interface \citep{Gilpin1980, TSW2015_EPL, TSW_PRL2017}. Recent high-resolution simulations of thermal convection over fractal boundaries \citep{toppaladoddi2021} with the same spectral properties as that of ice-ocean interface \cite{Rothrock:1980} provide support for this attribution. 

Although measuring the ocean heat flux at different depths in the Arctic mixed layer is possible, it is still challenging to do this over long periods of time across the entire basin. Laboratory experiments \citep{Gilpin1980, bushuk2019}, idealized high-resolution simulations \citep{couston2021, toppaladoddi2021_ice}, and turbulence modelling \citep{mcphee_book}, however, can be used to bridge this gap and construct a more complete picture of the spatio-temporal variability of the ocean heat flux. Another, comparatively less explored, approach is to describe the turbulent flow as a stochastic dynamical system and construct ordinary stochastic differential equations (SDEs) for velocity and temperature and solve them to obtain a model for the ocean heat flux. In the past, SDEs have been used to study the velocity field in spatially homogeneous isotropic turbulence \cite{pope2001, pope2002} and in parametrizations in climate models \citep{palmer2019}, but, to our knowledge, they have not been used to study stratified turbulent flows relevant to the Arctic mixed layer. In this study, we present a system of coupled ordinary SDEs for the velocity and temperature fluctuations and use it to study the heat flux in the Arctic mixed layer.

The following is a brief outline of the remainder of the paper. In \S\ref{section:model} we present the model and discuss how the parameters can be obtained from observational data. The velocity and temperature data from the Surface Heat Budget of the Arctic (SHEBA) project is briefly discussed in \S\ref{section:data}. Results and detailed comparisons with the SHEBA data are presented in \S\ref{section:results}, and conclusions are provided in \S\ref{section:conclusions}.

\section{The stochastic model} \label{section:model}
\subsection{Governing equations}
To obtain our phenomenological model for the fluctuations, we make use of the Reynolds-averaged description of turbulence \citep{Lumley} and decompose the velocity and temperature into mean and fluctuations about the mean. We assume there is no mean flow in the vertical direction and the vertical temperature profile is locally linear, which implies the mean temperature gradient is constant. The time evolution of the vertical velocity fluctuation, $\hat{w}$, and temperature fluctuation, $\hat{\theta}$, are modelled using the following dynamically-motivated stochastic differential equations:
\be
\frac{d\hat{w}}{d\hat{t}} = -\gamma_1 \, \hat{w} + g \, \alpha \, \hat{\theta} + b_1 \, \hat{\xi}_1(\hat{t}),
\label{eqn:velocity}
\ee
and
\be
\frac{d\hat{\theta}}{d\hat{t}} = -\gamma_2 \, \hat{\theta} - \beta \, \hat{w} + b_2 \, \hat{\xi}_2(\hat{t}).
\label{eqn:temp}
\ee
Here, $\hat{t}$ is the time, $\gamma_1$ and $\gamma_2$ are the relaxation frequencies of the velocity and temperature fluctuations, respectively; $g$ is acceleration due to gravity; $\alpha$ is the thermal expansion coefficient of seawater; $\beta$ is the mean temperature gradient in the upper layer of the ocean; $b_1$ and $b_2$ are the amplitudes of the noise terms; and $\hat{\xi}_1$ and $\hat{\xi}_2$ are Gaussian white noise terms with the property
\be
\left<\hat{\xi}_i(t_1) \, \hat{\xi}_j(t_2)\right> = \delta(t_1 - t_2) \, \delta_{ij}; \hspace{0.25cm} i, j = 1, 2,
\label{eqn:noise}
\ee
where $\delta(t_1 - t_2)$ is the Dirac delta function and $\delta_{ij} = 1$ if $i=j$ and $0$ otherwise. The angular brackets in equation \ref{eqn:noise} denote an ensemble average (i.e., over many statistical realizations, or a time average if the system is assumed to be ergodic, as done here). The instantaneous oceanic heat flux is given by
\be
F_w(\hat{t}) = \textcolor{blue}{\rho} \, C_p \, \hat{w}(\hat{t}) \, \hat{\theta}(\hat{t}),
\label{eqn:flux}
\ee
where $\rho = 1025$ kg \, m$^{-3}$ is the density of seawater and $C_p$ = 3985 J \, kg$^{-1}$ \, K$^{-1}$ is the specific heat of seawater, and the conductive heat flux has been assumed negligible outside of molecular boundary layers. 

Equations \ref{eqn:velocity} and \ref{eqn:temp} are heuristic simplified versions of the Boussinesq equations \cite{chandra2013} for momentum and heat balance, where the net effect of viscous or thermal diffusion and chaotic nonlinear fluctuations are represented by a linear relaxation term (first terms on right hand sides of equations \ref{eqn:velocity} and \ref{eqn:temp}) and a stochastic noise forcing (final terms in equations \ref{eqn:velocity} and \ref{eqn:temp}), respectively. The form of the relaxation terms is motivated by two reasons: first, that the dimensions of $\mathcal{D} \, \nabla^2$, where $\mathcal{D}$ is either momentum or thermal diffusivity, is that of the inverse of a time-scale \citep{young1994}; and second, the viscous term acts to damp the small-scale features in a turbulent flow. The second term on the right hand side of equation \ref{eqn:velocity} describes the coupling due to the thermal contribution to the buoyancy force. We assume that the buoyancy effects from changes in salinity can be subsumed into the stochastic forcing of velocity in equation \ref{eqn:velocity}. Hence, a separate equation for salinity fluctuations is not required in the model. The second term on the right hand side of equation \ref{eqn:temp} describes temperature changes due to advection against the background temperature gradient.
  
To non-dimensionalize equations \ref{eqn:velocity} and \ref{eqn:temp}, we choose velocity and temperature scales $w_0$ and $\theta_0$ as the standard deviations of the respective time series and $\gamma_1^{-1}$ as the time scale. Using these in the above equations and dropping the hats for dimensionless variables, we can express the equations in matrix form as
\be
\frac{d}{dt}\begin{bmatrix} w \\ \theta \end{bmatrix} = - \begin{bmatrix} 1 & -\Lambda_1 \\ \Lambda_2 & \Gamma \end{bmatrix} \, \begin{bmatrix} w \\ \theta \end{bmatrix} + \begin{bmatrix} B_1 & 0 \\ 0 & B_2 \end{bmatrix} \, \begin{bmatrix} \xi_1 \\  \xi_2 \end{bmatrix}.
\label{eqn:matrix}
\ee
The dimensionless parameters in equation \ref{eqn:matrix} are
\be
\Lambda_1 = \frac{g \, \alpha \, \theta_0}{w_0 \, \gamma_1}; \quad \Gamma = \frac{\gamma_2}{\gamma_1}; \quad \Lambda_2 = \frac{\beta \, w_0}{\theta_0 \, \gamma_1};
\nonumber
\ee

\be
B_1 = \frac{b_1}{w_0 \, \sqrt{\gamma_1}} \quad \mbox{and} \quad B_2 = \frac{b_2}{\theta_0 \, \sqrt{\gamma_1}}.
\ee
Our goal now is to develop solutions to equation \ref{eqn:matrix}. Before that, we determine the unknown dimensionless parameters from the given time series data.

\subsection{Determining the dimensionless parameters}
At polar latitudes the thermal expansion coefficient and thermal contribution to buoyancy is often comparatively small \citep{carmack2007}, and so when estimating the parameters we look for solutions with $\Lambda_1 \ll 1$. This implies that we are assuming the temperature to be a passive scalar. The veracity of this assumption will be put to test when we compare our results with observations.

Setting $\Lambda_1 = 0$, the eigenvalues of the coefficient matrix multiplying $w$ and $\theta$ in equation \ref{eqn:matrix} are $1$ and $\Gamma$. To determine $\Gamma$, we calculate $\gamma_1$ and $\gamma_2$ from the autocorrelation functions of $\hat{w}(\hat{t})$ and $\hat{\theta}(\hat{t})$ from the observational time series (Appendix \ref{sec:parameters}).


The remaining parameters are determined using some mathematical identities that are derived in Appendix \ref{sec:parameters}. With $\Lambda_1\rightarrow 0$, the equation for $w$ decouples, and hence we can sequentially solve (5) for $w(t)$ and $\theta(t)$. We can then calculate $\left<w(t)^2\right> = B_1^2/2$ (Appendix \ref{sec:parameters}). Noting that $\left<w^2\right> = 1$ due to the choice of scales for non-dimensionalization, we find $B_1 = \sqrt{2}$. A similar calculation for $\theta$ yields 
\be
B_2 = \sqrt{2 \, \Gamma - \frac{2 \, \Lambda_2^2}{1 + \Gamma}}.
\ee
Lastly, the value of $\Lambda_2$ can be determined by calculating the covariance of $w$ and $\theta$ (Appendix \ref{sec:parameters}), giving
\be
\Lambda_2 = - (1+\Gamma) \, \left<w \, \theta\right>.
\label{eqn:lambda2}
\ee
Hence, $\Lambda_2$ is determined from the covariance of the scaled velocity and temperature time series. Re-dimensionalizing equation \ref{eqn:lambda2} and rearranging leads to 
\be
\left<F_w\right> = \frac{\rho \, C_p \, \beta \, w_0^2}{\gamma_1 + \gamma_2},
\ee
which can be physically interpreted as follows. Fluctuations that transport significant heat feature a correlation of velocity and temperature anomalies, which persist over a timescale $\tau \sim (\tau_w^{-1} + \tau_{\theta}^{-1})^{-1}$ given by the harmonic mean of the damping timescales $\tau_w=\gamma_1^{-1}$ for velocity and $\tau_\theta=\gamma_2^{-1}$ for temperature. The persistence time scale, $\tau$, is thus dominated by the shorter damping timescale when there is a separation of damping timescales. The net heat accumulated by advection against the mean temperature gradient over this time is $\rho c_p w_0 \beta \tau$ which is transported with speed $w_0$ leading to the given expression for the heat flux. Note that this expression  recovers a classical bulk flux formula with heat flux proportional to mean velocity $U$ and temperature difference $\Delta T$ if $w_0 \propto U$ and $\beta \, w_0 \, \tau \propto \Delta T$ as might be expected for a turbulent shear flow with a background temperature gradient.

The above determines all the dimensionless parameters in terms of the statistical quantities, which can be obtained from the observed time series of velocity and temperature.

\subsection{Details of the numerical scheme}
After the dimensionless parameters are determined, the system \ref{eqn:matrix} is solved numerically using the Euler-Maruyama scheme \citep{higham2001}. The time step chosen for integration is $\Delta t = 10^{-3}$, and the equations are integrated for $10^5$ dimensionless time units, which is sufficiently long to obtain converged statistics. Both $w$ and $\theta$ are initialized using random numbers that are normally distributed with zero mean and unit variance.

\section{Data} \label{section:data}
We use the vertical velocity and temperature data from the Ocean Turbulence Mast Project conducted during the SHEBA expedition \citep{sheba}. The data were collected in the boundary layer underneath a drifting ice floe in the Arctic from October 1997 to September 1998. The clusters were spaced at a distance 4 m on the mast, and the uppermost cluster was at either a depth 4 m or 2 m below the ice-ocean interface (changed after the ice camp was moved following floe breakup in March 1998). Further details are provided along with the data by McPhee \citep{mcphee_sheba_data}. All the data used in our study are from the uppermost cluster.

The SHEBA data is partitioned into 15-minute segments, with the sampling frequency of either 0.5 or 1 Hz depending on the month. In order to understand the fluctuations on different timescales, the longest continuous interval of data was chosen for every month. The length of data differs from month to month, and ranges from about 7 hours in January, February, and December to 17.25 hours in September. The time series for $\hat{w}$ was taken as is, but to account for slow drift in the mean, the mean values of the temperature in each individual 15-min segments were subtracted from the temperature data before combining them. This results in a negligible trend in the time series for $\hat{\theta}$. The heat flux time series was then constructed using equation \ref{eqn:flux}.

In the following, we discuss the model results and their comparison with the SHEBA data.

\section{Results} \label{section:results}
\subsection{Relaxation times and mean temperature gradient}
\begin{figure}
\centering
\includegraphics[trim = 0 0 0 0, clip, width = 1\linewidth]{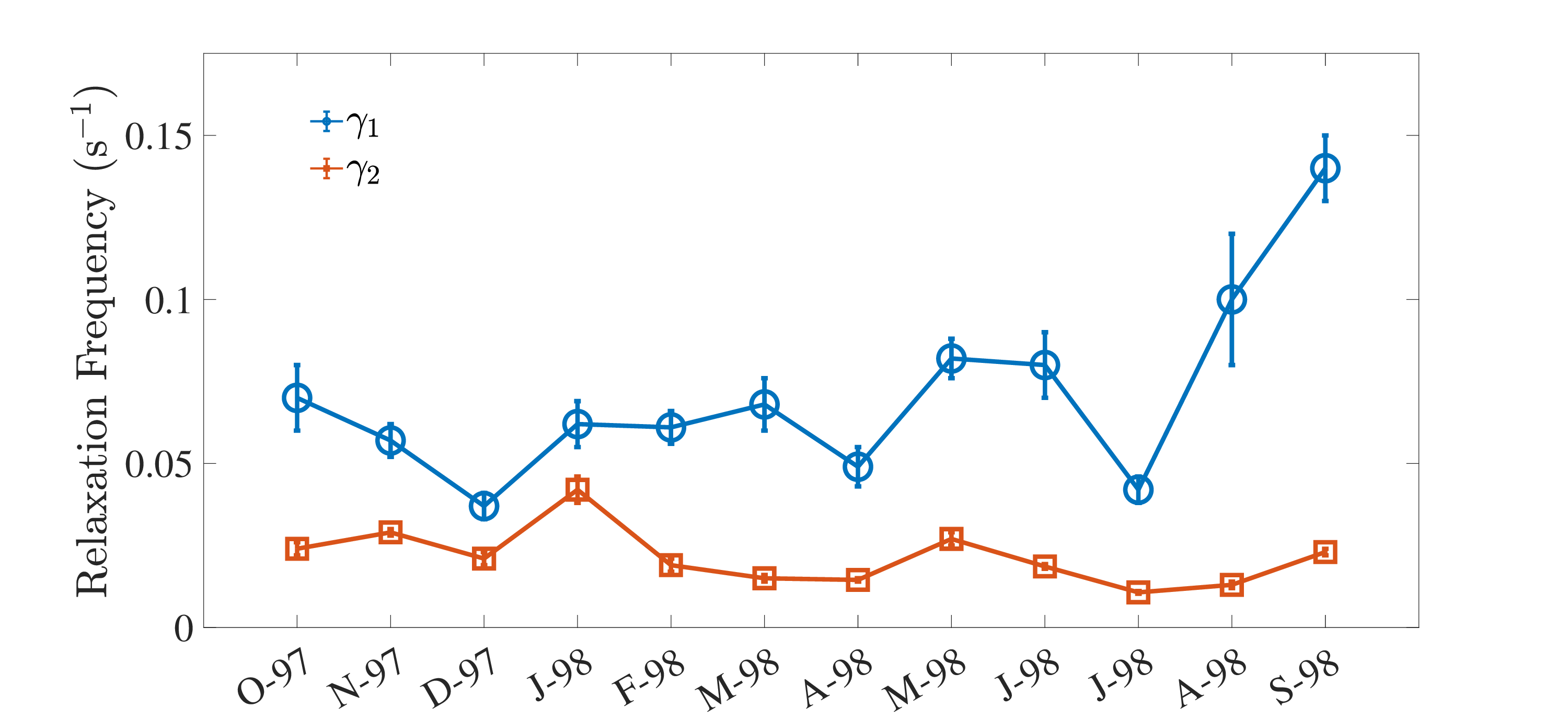}
\caption{Relaxation frequencies $\gamma_1$ and $\gamma_2$ for the different different months starting from October 1997 to September 1998. The error bars represent the 95$\%$ confidence intervals from the fits.}
\label{fig:relaxation}
\end{figure}

In figure \ref{fig:relaxation}, the monthly values of the relaxation frequencies $\gamma_1$ and $\gamma_2$ are shown. 
There are two observations that can be made from figure \ref{fig:relaxation}. First, the relaxation frequencies are not constants, but vary from month to month. They do however remain of a comparable order of magnitude throughout the time series, suggesting some consistency in the processes that control the dissipation of fluctuations. Secondly, the temperature fluctuations relax on slightly longer time scales than the velocity fluctuations. This implies that $\gamma_2 < \gamma_1$ and $\Gamma < 1$ for all months. The parameter $1/\Gamma$ is akin to the turbulent Prandtl number, which measures the ratio of turbulent diffusion of momentum and thermal anomalies and is $O(1)$ for all months. The relaxation frequencies might be related to the properties of the turbulent flow, such as the mean dissipation rate of the kinetic energy (see \S\ref{section:conclusions}), which vary from month to month, and hence the variation in $\gamma_1$ and $\gamma_2$.

\begin{figure}
\centering
\includegraphics[trim = 0 0 0 0, clip, width = 1\linewidth]{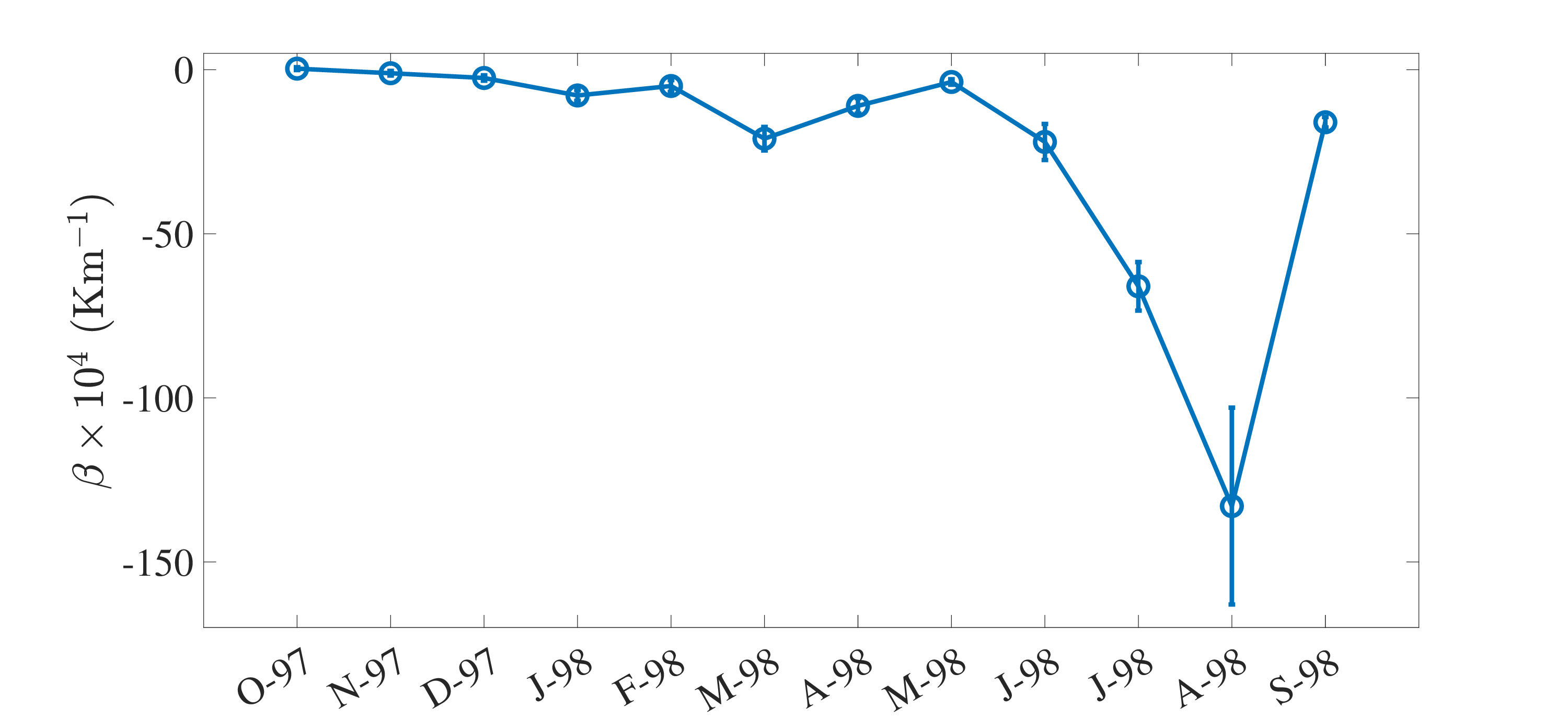}
\caption{Mean temperature gradient $\beta$ for the different months starting from October 1997 to September 1998. The error bars shown in the plot were calculated by propagating the systematic and statistical errors in velocity and temperature measurements.}
\label{fig:beta}
\end{figure}

We use the values of $\Gamma$ in equation \ref{eqn:lambda2} to calculate $\Lambda_2$ and, in turn, calculate $\beta$. This is shown in figure \ref{fig:beta}.
Except for the summer months of June, July and August, $|\beta| = O(10^{-5})$ - $O(10^{-3})$ K m$^{-1}$, indicating very weak temperature gradients in the mixed layer. The sign of $\beta$, except for the month of October, also indicates that the temperature increases with depth. This is consistent with the temperature of the ice-ocean interface being at the local freezing point, with warmer fluid below due to sources of heat at the base of the mixed layer \cite{timmermans2015}. The values of $\beta$ for the summer months are negative and larger than the values for the other months by one to three orders of magnitude. This might be due to thinner summer ice with lower concentration allowing enhanced absorption of solar radiation in the mixed layer, and increasing the temperature difference to the ice ocean interface which lies at the melting temperature. A further potential factor is the summer release of fresh meltwater which increases the density stratification, restricting the depth over which absorbed solar heating can be mixed, and thus increasing the magnitude of the temperature gradient. The data for October has $\beta > 0$ so that the temperature decreases with depth, which is counter intuitive. This could potentially be related to convective brine rejection during rapid initial ice growth in fall, injecting cold and saline water at the base of the mixed layer \citep[cf.][]{timmermans2015}. It should be emphasized here that the values of $\beta$ are calculated implicitly from single-point measurements, and hence may reflect very localized trends. A more complete picture might be obtained by analyzing vertical temperature profiles in the mixed layer, which is not possible with the available data set.

\subsection{Probability density functions}
\begin{figure}
\centering
\includegraphics[trim = 100 0 100 0, clip, width = 1\linewidth]{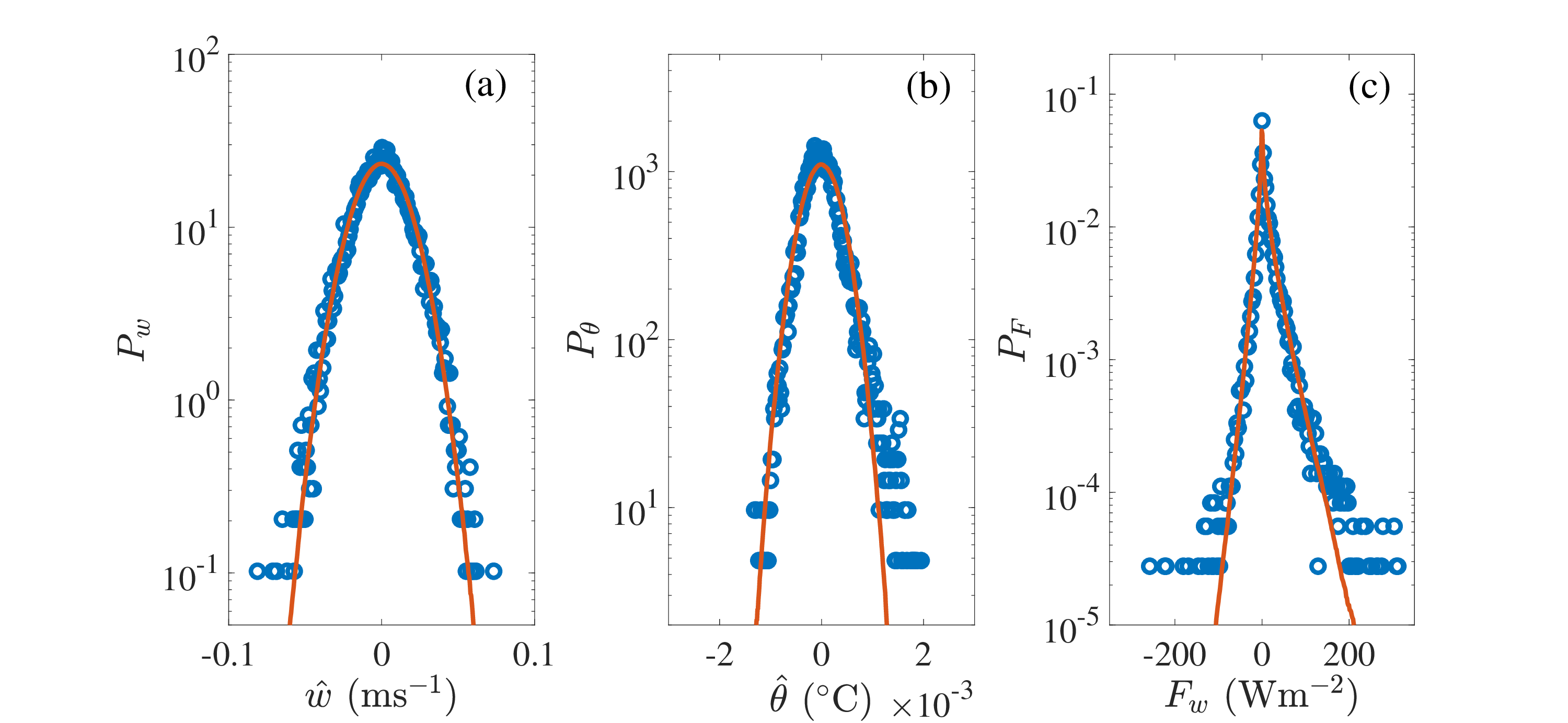}
\caption{PDFs for (a) vertical velocity ($\hat{w}$), (b) temperature ($\hat{\theta}$), and (c) heat flux ($F_w$) for the month of January. Circles denote PDFs from observations and the solid lines denote PDFs from the stochastic model. The PDFs for the observational and model data are generated using 200 and 400 bins, respectively.}
\label{fig:Jan_PDF}
\end{figure}

\begin{figure}
\centering
\includegraphics[trim = 100 0 100 0, clip, width = 1\linewidth]{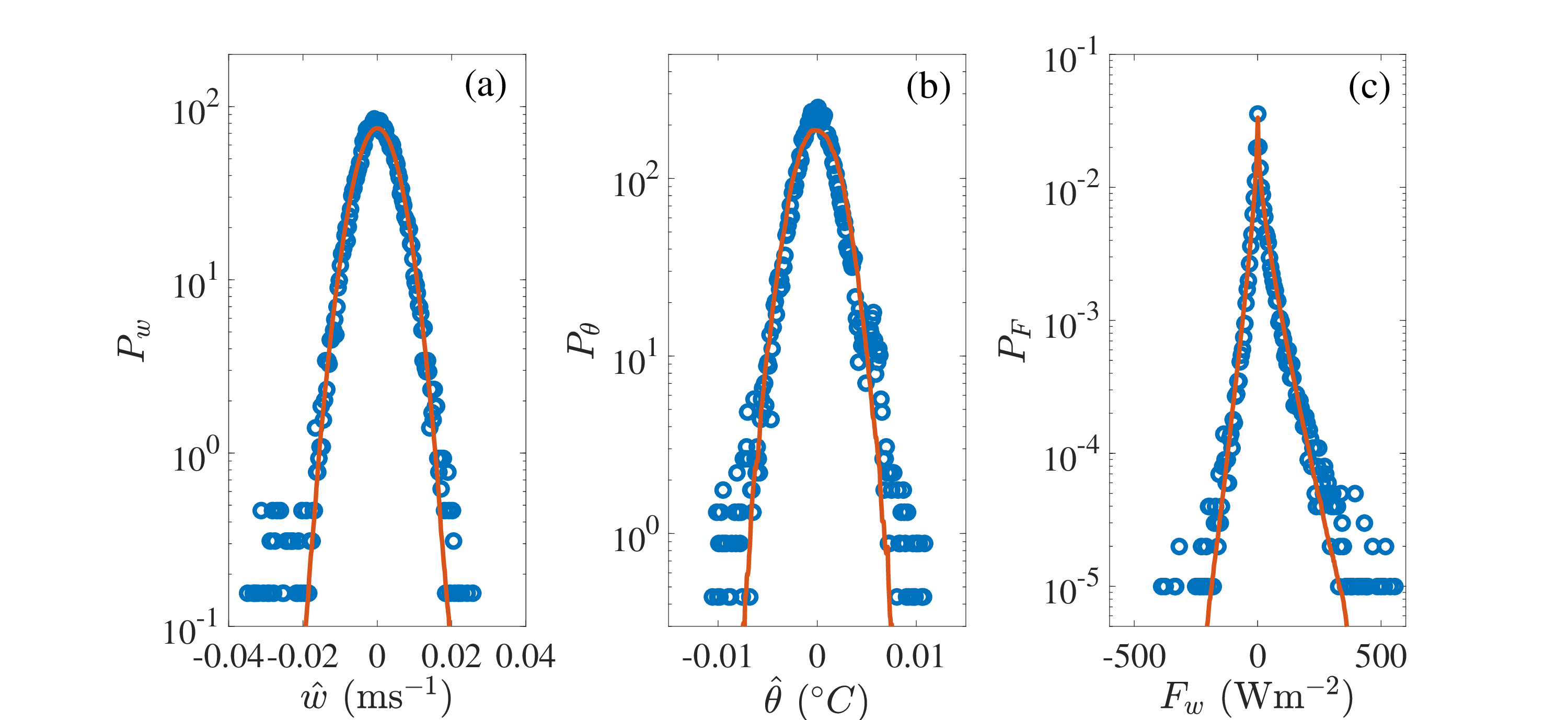}
\caption{PDFs for (a) vertical velocity ($\hat{w}$), (b) temperature ($\hat{\theta}$), and (c) heat flux ($F_w$) for the month of August. Circles denote PDFs from observations and the solid lines denote PDFs from the stochastic model. The PDFs for the observational and model data are generated using 200 and 400 bins, respectively.}
\label{fig:Aug_PDF}
\end{figure}

\begin{figure}
\centering
\includegraphics[trim = 100 0 100 0, clip, width = 1\linewidth]{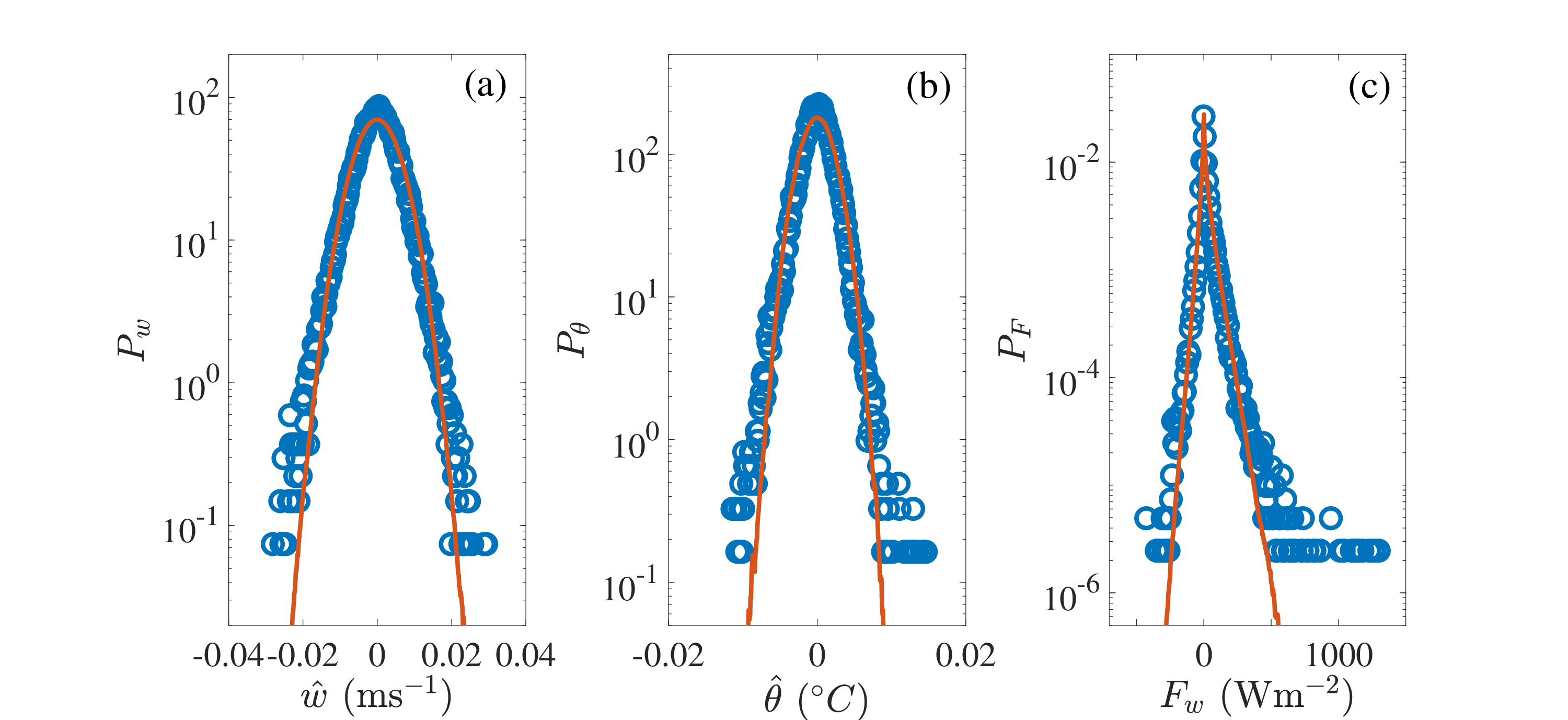}
\caption{PDFs for (a) vertical velocity ($\hat{w}$), (b) temperature ($\hat{\theta}$), and (c) heat flux ($F_w$) for the month of July. Circles denote PDFs from observations and the solid lines denote PDFs from the stochastic model. The PDFs for the observational and model data are generated using 200 and 400 bins, respectively.}
\label{fig:Jul_PDF}
\end{figure}

In figure \ref{fig:Jan_PDF}, the PDFs for $\hat{w}$, $\hat{\theta}$, and $F_w$ -- which are denoted by $P_w$, $P_{\theta}$, and $P_F$, respectively -- are shown for the month of January on semi-log plots. 
It is apparent from figure \ref{fig:Jan_PDF}(a) that the velocity fluctuations are described well by the Langevin equation (top row of equation \ref{eqn:matrix} with $\Lambda_1 = 0$). This implies that the assumption $\Lambda_1 \ll 1$, or in other words that the temperature is a passive scalar, is observationally consistent. Qualitatively similar behaviour is observed for the other months as well (e.g., figures \ref{fig:Aug_PDF} and \ref{fig:Jul_PDF}), thus indicating that $\Lambda_1 \ll 1$ is valid for the whole dataset. Although the PDF for $\hat{w}$ from observations is not exactly a Gaussian (its skewness and kurtosis are $\approx -0.02$ and $3.32$, respectively versus $0$ and $3$ expected for a Gaussian), it is still described well by the model curve which is a Gaussian. The near Gaussian behaviour of the velocity fluctuations is often observed in homogeneous turbulent flows \citep{batchelor1953, townsend1980}, and it is interesting that we here have similar behaviour in the under ice boundary layer which likely experiences shear and buoyancy forces.

In contrast to the velocity fluctuations, there is a marked departure from Gaussianity in the temperature fluctuations. The temperature PDF from the model is approximately Gaussian, but the PDF from observations is not. This is seen in figure \ref{fig:Jan_PDF}(b), where there is a clear difference between the tails of the PDFs from the model and observation. However, the model still overall describes the observations satisfactorily well. Note that the discrete quantization of probability values seen in the tails of the observed distributions is indicative of finite sample size effects, with only one, two, three etc occurrences in each bin for each quantized probability level. Thus the observed values in the tails carry greater uncertainty as an estimator of the true probability distribution. 

Figure \ref{fig:Jan_PDF}(c) shows that the PDF of heat flux is clearly non-Gaussian and skewed, and approximately consistent with two patched stretched exponentials, as previously observed by \citet{mcphee1992}. (See Appendix \ref{sec:nonGauss} where we derive an explicit expression for $P_F$ for correlated Gaussian distributions of $\hat{w}$ and $\hat{\theta}$, which predicts $P_F$ has asymmetric exponential tails modified by a power-law pre-factor.) The PDF from the model is able to describe the observations well, except in the tails. These rare events have large instantaneous heat flux values associated with them as described by \citet{mcphee1992}, but the model in its current form does not capture these. The source of these large fluctuations is likely the large non-Gaussian fluctuations in the vertical velocity and temperature, which, as described earlier, are not captured by the model.

In figure \ref{fig:Aug_PDF}, we show the PDFs for the month of August. As can be seen from figures \ref{fig:Aug_PDF}(a) -- \ref{fig:Aug_PDF}(c), the model PDFs describe the observations well, with the exception of tails.
For the month of July too (figure \ref{fig:Jul_PDF}), we observe large differences in the tails between the model and observations.
It is apparent that the model is less effective in capturing the tails in the temperature PDF. As a result of this, there are differences between the tails of the model and observational PDFs of heat flux as well. This pattern of behaviour is qualitatively similar in the other months.

One possible reason for the difference in the results from the model and observations is that the nature of the noise in the temperature equation might not be Gaussian. There might be physical processes that cause large fluctuations more often than what would be expected for Gaussian processes, but it is not apparent what these processes might be and what the time-scales associated with them are. Another reason might be that the noise terms in the equations of motion are not additive, but are multiplicative. In this case, it is unclear how to constrain the functional forms of the noise amplitudes from the limited data. Constructing a separate stochastic model to test these hypotheses is beyond the scope of the current work.

\begin{figure}
\centering
\includegraphics[trim = 75 0 100 0, clip, width = 1\linewidth]{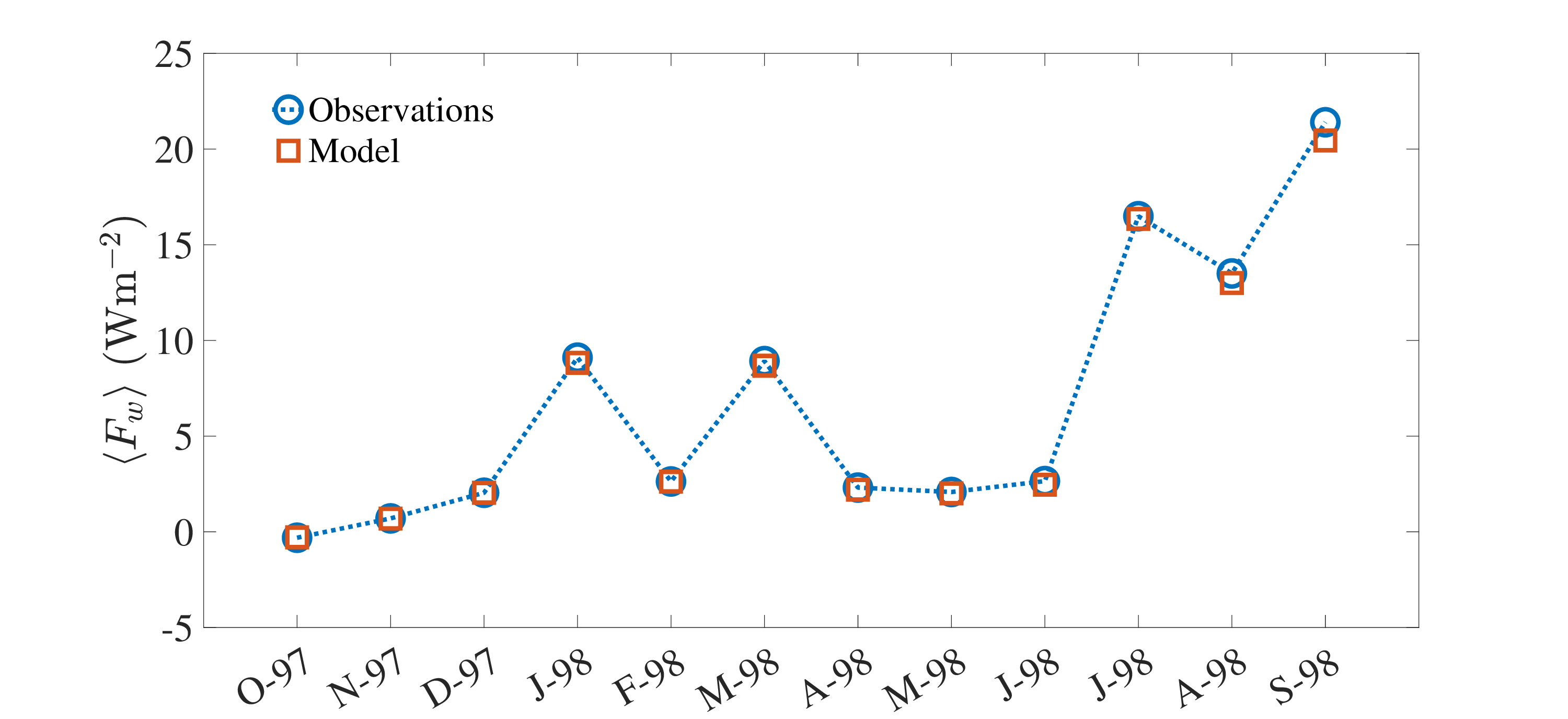}
\caption{Comparison between the mean heat values from the model and observations for the different months. }
\label{fig:mean_heat_flux}
\end{figure}

However, we recall that choosing model parameter values according to equation \ref{eqn:lambda2} results in the model accurately capturing the mean value of the heat flux for a sufficiently long time series. In figure \ref{fig:mean_heat_flux}, we compare the mean heat flux values obtained from the model and from observations, with the close agreement indicating the time series are sufficiently long to adequately characterize the model properties using  the statistically averaged relation \ref{eqn:lambda2}. The fact that the mean heat flux (figure \ref{fig:mean_heat_flux}) and near-peak structure of the pdfs (figures \ref{fig:Jan_PDF}, \ref{fig:Aug_PDF}, and \ref{fig:Jul_PDF}) are simultaneously well approximated suggests that the deviations in the tail of the pdfs do not contribute significantly to the average heat flux. Whilst these events in the tails have large magnitude, they are very rare. The mean heat flux is instead controlled by the correlation of $w$ and $\theta$ via advection against the mean temperature gradient, which induces a skewness to the central core of the PDF. It should be noted here that although the mean temperature gradient, $\beta$, is weak (figure \ref{fig:beta}) for most of the year, it still has a controlling effect on the dynamics of the system as the velocity and temperature fluctuations are coupled through this term. This coupling ultimately leads to the correct shape of the PDFs for $F_w$ in our model.

\section{Conclusions} \label{section:conclusions}
The following are the main conclusions from our study.
\begin{enumerate}

\item We have developed an observationally consistent stochastic model to describe fluctuations in the vertical velocity, temperature, and heat flux in the Arctic mixed layer. The dimensionless parameters in the model are determined using correlation and cross-correlation functions of the temperature and velocity time series from observations.

\item We showed that by assuming the thermal contribution to the buoyancy term in the equation for vertical velocity to be small, we were able recover the observed PDF for $w$, which is approximately Gaussian. This indicates that to leading order, the temperature in the Arctic mixed layer can be treated as a passive scalar.

\item The temperature and heat flux PDFs from our model are in good overall agreement with the ones from observations.

\item The theory, in its current form, requires certain averaged quantities from the observations as input parameters. However, these quantities are obtained from different aspects of the observations; the resulting agreement between the theory and observations (figures \ref{fig:Jan_PDF}--\ref{fig:mean_heat_flux}) shows that the model has the appropriate mathematical and physical structure to produce observationally consistent statistics. We should emphasize here that we only use first and second order moments of the temperature and velocity time series to determine the dimensionless parameters in the model, but nevertheless obtain good agreement for the entire PDFs.

\item A shortcoming of the model is that it does not capture the rare events (tails of the PDFs) in temperature and heat flux. However, this has negligible effect on the mean values of the heat flux (figure \ref{fig:mean_heat_flux}).

\end{enumerate}

We showed that provided the key parameters are obtained from the observational time series, our stochastic model can be used to obtain reliable statistics of the heat flux. However, some of these parameters  appear to vary seasonally with the ocean conditions. For use as a prognostic parameterization in regional sea ice simulations it is necessary to relate the parameters to coarse grained variables that could be described or inferred in the regional simulations. Hence, these parameters would have to be estimated from other bulk quantities.

One might try to relate the relaxation frequencies $\gamma_1$ and $\gamma_2$ to molecular diffusivities as $\gamma_1 = C_1 \, \nu \, k^2$ and $\gamma_2 = C_2 \, \kappa \, k^2$, where $\nu$ and $\kappa$ are the kinematic viscosity and thermal diffusivity, respectively, $k=2\pi/L$ is the dominant wavenumber for the characteristic length scale $L$, and $C_1$ and $C_2$ are dimensionless constants of $O(1)$. These expressions pose both conceptual and practical difficulties. The key question to be addressed to make any progress is: what is the characteristic length scale $L$? Two potential choices for this length scale are the Taylor microscale ($\lambda_T$) and the Kolmogorov scale ($\eta$) defined as $\lambda_T = \sqrt{15 \nu/\epsilon} \, v_{rms}$ and $\eta = \left(\nu^3/\epsilon\right)^{1/4}$, where $\rho \, \epsilon$ is the mean dissipation rate of the fluid kinetic energy and $v_{rms}$ is the root-mean-square of velocity fluctuations. We estimate $\epsilon$ in terms of the work done by shear stresses per unit volume per unit time in the mixed layer using a bulk formula for the shear stress. This gives $\rho \, \epsilon \approx \tau \, U/H$ = $ \rho \, C_d \, U^3/H$, where $\tau$ is the shear stress, $C_d$ is the drag coefficient, $U$ is the mean relative ice-ocean velocity along the horizontal, and $H$ is the depth of the mixed layer. We estimate $v_{rms} \approx 0.01$ m/s by order of magnitude based on the observational data. As an order of magnitude estimate we use $U\approx 0.1$ m/s, $C_d = 5 \times 10^{-3}$, $\nu=2\times 10^{-6}$ m$^2$/s and $H \approx 10 $m to estimate  $\eta \approx 0.002$ m and $\lambda_T \approx 0.08$ m. Setting $L \sim \eta$ based off the Kolmogorov scale yields $\gamma_1 \approx 20$ s$^{-1}$, which is much larger than the observed frequencies in figure \ref{fig:relaxation}. However, using the Taylor microscale $L \sim \lambda_T$ results in $\gamma_1 \approx 0.013$ s$^{-1}$ which is intriguingly of similar order of magnitude to the observed velocity and thermal dissipation frequency scales in figure \ref{fig:relaxation}. 

Future work might consider a detailed analysis of observational or numerical simulation data to evaluate this hypothesis more carefully, and understand why the thermal dissipation frequency scale does not exactly vary in proportion to the velocity dissipation frequency scale. We should also note here that both $U$ and $H$ vary with seasons, which potentially contributes to the seasonal variations in $\gamma_1$ and $\gamma_2$ at the leading order. Hence, a systematic inclusion of these temporal variations in the future work is also necessary. Alternatively, data from year-round, high-resolution measurements of velocity and temperature profiles would permit accurate calculations of the gradients, which in turn will lead to more accurate estimates of kinetic and thermal dissipation rates and hence $\gamma_1$ and $\gamma_2$. Furthermore, the high-resolution data would also permit the calculation of more accurate values of $\beta$.

In the absence of such data, one of the following two approaches could be taken to determine $\Lambda_2$. The first one might determine a relationship between the mean temperature gradient $\beta$ and other coarse grained variables using observational data. Because of the large spacing between the clusters (4 m), a reliable temperature gradient cannot be calculated from the SHEBA data. However, high resolution vertical profiles of temperature are now available from Ice Tethered Profiler (ITP) measurements in the different regions of the Arctic \citep{toole2011}. The second method is that one could use the bulk relations typically used to predict mean heat fluxes to estimate the covariance $\left<w \theta\right>$ \citep{mcphee_book}, which can then be used to calculate $\Lambda_2$ using equation \ref{eqn:lambda2}. One also needs to estimate the standard deviations of velocity and temperature fluctuations, with possible candidate scalings proportional to the mean horizontal velocity within the mixed layer and temperature difference between ice-ocean interface and mixed layer (in line with the dimensional underpinnings of typical bulk flux formulae).


The value of our method is in that it provides a way to obtain the observationally consistent probability density functions of the ocean heat flux from knowing only certain bulk quantities. This may be helpful in calculating growth rates of sea ice in both regional and global climate models with a sea-ice component in them, provided that the key model parameters are known. 
\\~\\
\section*{Acknowledgements}
The authors thank one of the anonymous referees for suggesting the analysis presented in Appendix \ref{sec:nonGauss}.

\appendix
\section{Dimensionless parameters} \label{sec:parameters}

To obtain the relaxation frequencies, $\gamma_1$ and $\gamma_2$, we first calculate the autocorrelation functions $C_w$ and $C_{\theta}$, which are defined by
\be
C_w(\hat{t}) = \left<\hat{w}(\hat{t}_0) \, \hat{w}(\hat{t}_0 + t)\right> \quad \text{and} \quad C_{\theta}(\hat{t}) = \left<\hat{\theta}(\hat{t}_0) \, \hat{\theta}(\hat{t}_0 + t)\right>,
\ee
where the averages are computed via integration over time. The relaxation frequencies are then obtained by fitting exponential curves to the correlation functions.

To determine $B_1$, we solve the top row of equation \ref{eqn:matrix} using $\Lambda_1 = 0$ and an integrating factor. The solution is
\be
w(t) = w_0 \, e^{-t} + B_1 \, e^{-t} \, \int_0^t e^{t_1} \, \xi_1(t_1) \, dt_1,
\ee
where $w_0$ is the initial condition, which can be set to zero without the loss of generality because we are interested in the long time average ($t \rightarrow \infty$). The autocorrelation function for $w$ is 
\be
\left<w(t) \, w(s)\right> = B_1^2 \, e^{-(t+s)} \, \int_0^t \int_0^s e^{(t_1 + t_2)} \, \left<\xi_1(t_1) \, \xi_1(t_2)\right> dt_2 \, dt_1.
\ee
The integral can be evaluated using the property of white noise (equation \ref{eqn:noise}), and in the limits $t \rightarrow \infty$ and $s \rightarrow \infty$ with $|t-s|$ finite gives
\be
\left<w(t) \, w(s)\right>  = \frac{B_1^2}{2} \, e^{-|t-s|}.
\label{eqn:wauto}
\ee
Setting $t = s$ and noting that $\left<w^2\right> = 1$, we get $B_1 = \sqrt{2}$.

Next, in order to determine $\Lambda_2$ and $B_2$, we solve the bottom row of equation \ref{eqn:matrix} using an integrating factor to give
\begin{widetext}
\be
\theta(t) = \theta_0 \, e^{-\Gamma \, t} - \Lambda_2 \, e^{-\Gamma \, t} \, \int_0^t e^{\Gamma \, t_1} \, w(t_1) \, dt_1 + B_2 \, e^{-\Gamma \, t} \, \int_0^t e^{\Gamma \, t_1} \, \xi_2(t_1) \, dt_1.
\label{eqn:thetasol}
\ee
\end{widetext}
The initial condition $\theta_0$ can again be set to $0$ without any loss of generality. Multiplying equation \ref{eqn:thetasol} by $w(t)$, taking the ensemble average and noting that $\left<w(t) \, \xi_2(t_1)\right> = 0$ gives
\be
\left<w \, \theta \right> = - \Lambda_2 e^{-\Gamma \, t}  \, \int_0^t e^{\Gamma \, t_1} \, \left<w(t) \, w(t_1)\right> \, dt_1.
\ee
Using the result in equation \ref{eqn:wauto}, this can be solved to give
\be
\Lambda_2 = - (1+\Gamma) \, \left<w \, \theta\right>
\label{eqn:lambda2_appendix}
\ee
in the limit $t \rightarrow \infty$. 

Lastly, to calculate $B_2$, we calculate the variance of $\theta$ which is given by the expression
\begin{widetext}
\be
\left<\theta^2\right> = \Lambda_2^2 \, e^{-2 \, \Gamma \, t} \, \int_0^t \int_0^t e^{\Gamma \, (t_1 + t_2)} \, \left<w(t_1) \, w(t_2)\right> \, dt_2 \, dt_1 + B_2^2 \, e^{-2 \, \Gamma \, t} \, \int_0^t \int_0^t e^{\Gamma \, (t_1 + t_2)} \, \left<\xi_2(t_1) \, \xi_2(t_2)\right> \, dt_2 \, dt_1.
\ee
\end{widetext}
The integrals can be evaluated to give
\be
B_2 = \sqrt{2 \, \Gamma \, \left<\theta^2\right> - \frac{2 \, \Lambda_2^2}{1 + \Gamma}}
\ee
in the limit $t \rightarrow \infty$. Noting that $\left<\theta^2\right> = 1$, we finally get
\be
B_2 = \sqrt{2 \, \Gamma - \frac{2 \, \Lambda_2^2}{1 + \Gamma}}.
\ee

\section{Analytical model for the non-Gaussian distribution} \label{sec:nonGauss}
Here, we present an analysis that shows that the product of two Gaussian variables, which are not independent, produces a non-Gaussian variable. Using the result that the solution to a linear multi-variate Fokker-Planck equation is a Gaussian \citep{van1992}, we write the stationary joint distribution for $w$ and $\theta$ as
\be
P_{w,\theta}(w,\theta) = \mathcal{N} \, \exp(-\mathcal{A}\, w^2 - \mathcal{B} \, w \, \theta - \mathcal{C} \, \theta^2),
\label{eqn:jointPDF}
\ee
where $\mathcal{N}$, $\mathcal{A}$, $\mathcal{B}$, and $\mathcal{C}$ are determined by second-order moments of $P(w, \theta)$. The standard form of the two-dimensional Gaussian distribution is
\be
P_{w,\theta}(w,\theta) = \frac{1}{2 \, \pi \, \sqrt{|\boldsymbol{\Sigma}|}} \, \exp\left(-\frac{1}{2} \, \boldsymbol{X}^T \cdot \boldsymbol{\Sigma}^{-1} \cdot \boldsymbol{X}\right),
\label{eqn:jointPDF}
\ee
where
\be
    \boldsymbol{X} = \begin{bmatrix}
           w \\
           \theta \\
         \end{bmatrix},
\label{eqn:standardGauss}
\ee
$\boldsymbol{X}^T$ is the transpose of $\boldsymbol{X}$, $\boldsymbol{\Sigma}$ is the covariance matrix and $\boldsymbol{\Sigma}^{-1}$ and $|\boldsymbol{\Sigma}|$ are its inverse and determinant, respectively. The entries of covariance matrix are given by $\Sigma_{ij} = \left<X_i \, X_j\right>$, where $i, j = 1, 2$. Comparing equations \ref{eqn:jointPDF} and \ref{eqn:standardGauss}, it is straightforward to find that
\be
\boldsymbol{\Sigma}^{-1} = 
\begin{bmatrix}
    2 \, \mathcal{A}       & \mathcal{B}  \\
    \mathcal{B}       & 2 \, \mathcal{C} 
\end{bmatrix},
\ee
and hence
\be
\boldsymbol{\Sigma} = \frac{1}{4 \, \mathcal{A} \, \mathcal{C} - \mathcal{B}^2}
\begin{bmatrix}
    2 \, \mathcal{C}       & -\mathcal{B}  \\
    -\mathcal{B}       & 2 \, \mathcal{A} 
\end{bmatrix}.
\label{eqn:sigma}
\ee
Evaluating $\Sigma_{ij} = \left<X_i \, X_j\right>$ using equations \ref{eqn:standardGauss} and \ref{eqn:sigma} yields three simultaneous equations for $\mathcal{A}$, $\mathcal{B}$, and $\mathcal{C}$ with solutions:
\be
\mathcal{A} = \frac{\left<\theta^2\right>}{2 \left(\left<\theta^2\right> \, \left<w^2\right> - \left<w \, \theta\right>^2 \right)},
\ee

\be
\mathcal{B} = \frac{-\left<w \, \theta\right>}{\left(\left<\theta^2\right> \, \left<w^2\right> - \left<w \, \theta\right>^2 \right)},
\ee
and
\be
\mathcal{C} = \frac{\left<w^2\right>}{2 \left(\left<\theta^2\right> \, \left<w^2\right> - \left<w \, \theta\right>^2 \right)}.
\ee
The normalisation factor is then given by
\be
\mathcal{N} = \frac{\sqrt{\left(4 \, \mathcal{A} \, \mathcal{C} - \mathcal{B}^2\right)}}{2 \, \pi}.
\ee
\\
To find the PDF for the instantaneous heat flux we first note that a multivariate PDF, $P_{w, \theta}(w,\theta)$, for random variables $w$ and $\theta$ transforms as 
\be
P_{w, \theta}(w, \theta) \, dw \, d\theta = P_{w, \theta}\left[w(x,y),\theta(x,y)\right] \, \abs{\frac{\partial(w,\theta)}{\partial(x,y)}} \, dx \, dy,
\ee
which implies
\be
P_{x, y}(x, y) = P_{w, \theta}\left(w,\theta\right) \, \abs{\frac{\partial(w,\theta)}{\partial(x,y)}}.
\ee
If we choose $x = w$ and $y = F = w \, \theta$, where $F$ is the dimensionless heat flux, we get
\be
P_{w, F}(w, F) = \mathcal{N} \, \frac{1}{|w|} \, \exp\left(-\mathcal{A}\, w^2 - \mathcal{B} \, F - \mathcal{C} \, \frac{F^2}{w^2}\right).
\ee
The marginal PDF for the heat flux can now be found using
\be
P_F = \mathcal{N} \, \int_{-\infty}^{\infty} \, \frac{1}{|w|} \, \exp\left(-\mathcal{A}\, w^2 - \mathcal{B} \, F - \mathcal{C} \, \frac{F^2}{w^2}\right) \, dw, 
\ee
or
\be
P_F = 2 \, \mathcal{N} \, \int_0^{\infty} \, \frac{1}{|w|} \, \exp\left(-\mathcal{A}\, w^2 - \mathcal{B} \, F - \mathcal{C} \, \frac{F^2}{w^2}\right) \, dw
\ee
because the integrand is an even function of $w$. This integral is evaluated assuming $\mathcal{A}, \mathcal{C} > 0$ and substituting
\be
w = \left(F^2 \, \frac{\mathcal{C}}{\mathcal{A}}\right)^{1/4} \, \exp(z/2),
\ee
which gives
\be
P_F = \mathcal{N} \, \exp(-\mathcal{B} \, F) \, \int_{-\infty}^{\infty} \exp\left[-2 \, |F| \, \sqrt{\mathcal{A} \, \mathcal{C}} \, \cosh(z)\right] \, dz.
\ee
This can alternatively be expressed as
\be
P_F = 2 \, \mathcal{N} \, \exp(-\mathcal{B} \, F) \, K_0(-2 \, |F| \, \sqrt{\mathcal{A} \, \mathcal{C}}),
\label{eqn:analy_1}
\ee
using the integral definition of $K_0(X)$, the zeroth order modified Bessel function of second kind \citep{stegun}. For $|F| \gg 1$ we get the asymptotic behaviour of $P_F$ as (see \citep{stegun} for asymptotic behaviour of $K_0(X)$)
\be
P_F \sim \frac{\mathcal{N}}{\sqrt{|F|}} \, \exp(-\mathcal{B} \, F - 2 \, |F| \, \sqrt{\mathcal{A} \, \mathcal{C}}).
\label{eqn:analy_2}
\ee

\begin{figure}
\centering
\includegraphics[trim = 75 0 100 0, clip, width = 1\linewidth]{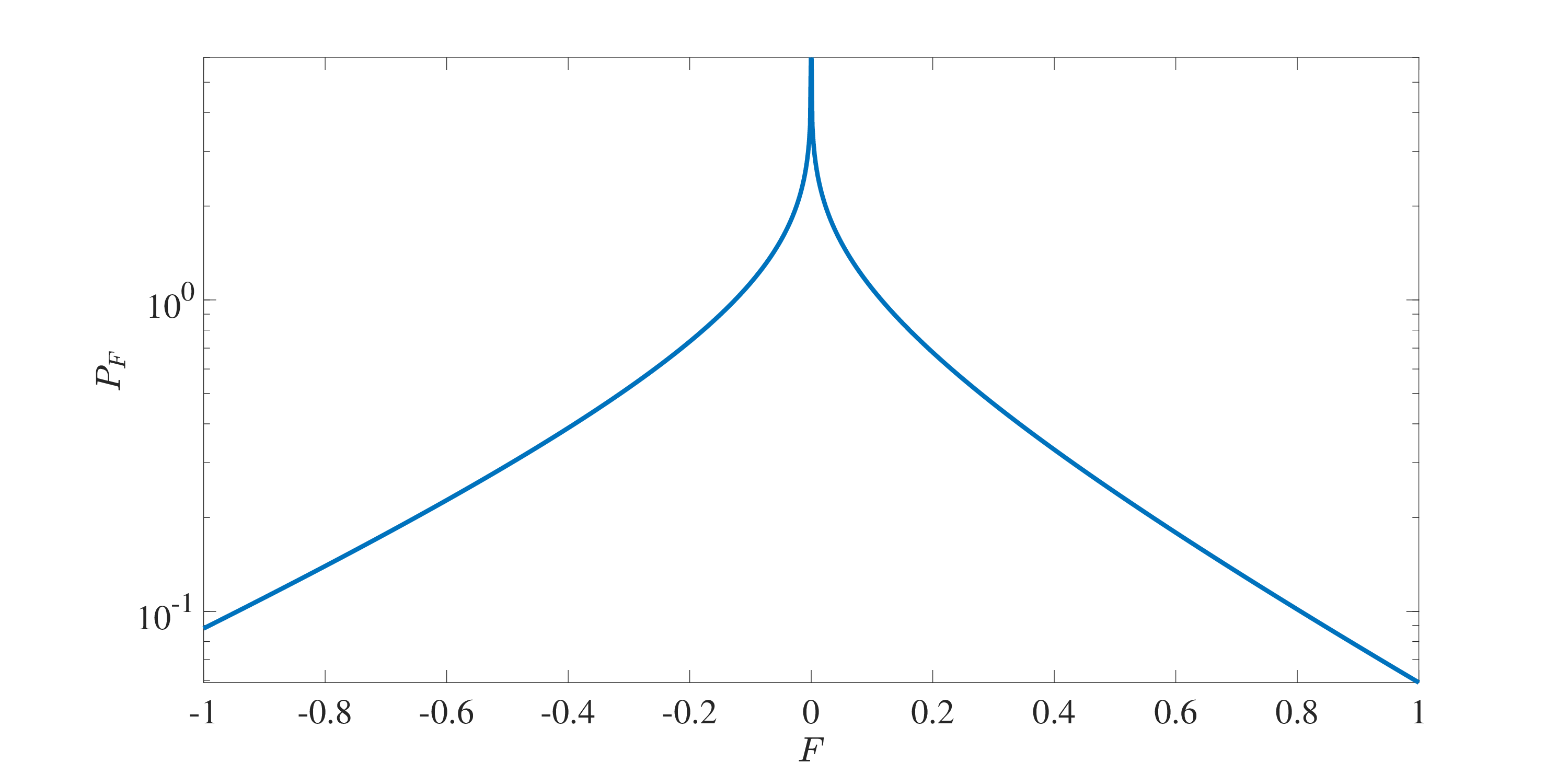}
\caption{Plot of $P_F$ (equation \ref{eqn:analy_1}) for $\mathcal{A} = 1$, $\mathcal{B} = 0.2$, and $\mathcal{C} = 1$. The singularity at $F = 0$ is masked due to finite resolution.}
\label{fig:analy}
\end{figure}

There are two key features of this analytical solution (see figure \ref{fig:analy}). 
First, $P_F$ has an integrable singularity at $F = 0$; and second, the PDF consists of two exponential tails, modified by power-law pre-factor, with different decay scales for $F <0$ and $F>0$, which implies $P_F$ is asymmetric. This asymmetry is in clear qualitative agreement with the observations shown in figures \ref{fig:Jan_PDF}-\ref{fig:Jul_PDF}. However, the two branches of the PDFs from observations are not exponentials, but stretched exponentials \citep{mcphee1992}. This discrepancy is because in obtaining the analytical solution we have assumed the PDFs of $w$ and $\theta$ are Gaussian; but the observations show that this is only approximately true.


\color{black}

\bibliography{jfm_gilpin} 
\end{document}